\documentclass[twocolumn,showpacs,prl]{revtex4}
\usepackage{graphicx}
\begin{document}

\title{One-dimensional Bose chemistry: effects of non-integrability}

\author{V. A. Yurovsky}

\author{A. Ben-Reuven}

\affiliation{School of Chemistry, Tel Aviv University, 69978 Tel Aviv,
Israel}

\author{M. Olshanii}

\affiliation{Department of Physics \& Astronomy, University of
 Southern
California, Los Angeles, CA 90089-0484, USA}

\date{\today}

\begin{abstract}Three-body collisions of ultracold
identical Bose atoms under tight cylindrical confinement are
analyzed. A Feshbach resonance in two-body collisions is
described by a two-channel zero-range interaction. Elimination
of the closed channel in the three-body problem reduces the
interaction to a one-channel zero-range one with an energy
dependent strength. The related problem with an energy
independent strength (the Lieb-Liniger-McGuire model) has an
exact solution and forbids all chemical processes, such as
three-atom association and diatom dissociation, as well as
reflection in atom-diatom collisions. The resonant case is
analyzed by a numerical solution of the Faddeev-Lovelace
equations. The results demonstrate that as the internal
symmetry of the Lieb-Liniger-McGuire model is lifted, the
reflection and chemical reactions become allowed and may be
observed in experiments.\end{abstract}

\maketitle

Ultracold atoms under tight cylindrical confinement can reach the
``single-mode'', or quasi-one-dimensional (1D) regime, where only the
ground state of transverse motion is significantly populated. Quasi-1D
atomic quantum gases have been realized recently in elongated atomic
 traps
(see  \cite{G01a,S02,K02,R03}), two-dimensional optical lattices (see
\cite{G01,W04,F05}), atomic waveguides (see \cite{L02} and references
therein), and atomic integrated optics devices (see  \cite{F02,S05}
 and
references therein). These systems  attract increasing attention due
 to
their possible applications to atomic interferometry, quantum
 measurement,
and  quantum computation.

Properties of quasi-1D  systems are frequently analyzed using  the
Lieb-Liniger-McGuire (LLMG) model \cite{LL63,McGuire64} with $\delta
 $ function
atom-atom interaction $U_{a}\delta \left( z\right) $, where $z$  is
 the interatomic distance. The
interaction strength $U_{a}$  has been related in Ref.\ \cite{O98}
(being denoted
as $g_{1D}$) to the three-dimensional (3D) elastic scattering length
 $a_{3D}$  and the
waveguide transverse frequency $\omega _{\perp }$. The LLMG model has
 an exact Bethe-ansatz
solution \cite{LL63,McGuire64} expressed as the superposition of
 plane waves
with all possible permutations of the asymptotic momenta $p_{j}$, one
 per each
atom. Therefore, the atoms can exchange their momenta, but the
 asymptotic
momentum set $\{p_{j}\}$ remains unchanged.

If the interatomic interaction is attractive ($U_{a}<0$), the atoms
can form a two-body (2B) bound state (diatom), described by complex
asymptotic momenta $p\pm i\kappa $, where $\kappa =-{m\over 2}U_{a}$
 serves as a reciprocal diatom
size and $m$ is the atomic mass. However, a three-body (3B)
 association
is forbidden in the LLMG model, as it corresponds to a change of two
real momenta in the set $\{p_{1},p_{2},p_{3}\}$ for a three-atom
 system by complex
ones in the set $\{-p,p/2+i\kappa ,p/2-i\kappa \}$. The dissociation
 and reflection
in atom-diatom collisions are forbidden by the same reason (the last
process corresponds to a transition $\{-p,p/2+i\kappa ,p/2-i\kappa
 \}\rightarrow \{p,-p/2+i\kappa ,-
p/2-i\kappa \}$). Therefore, all chemical processes, i. e.,
 atom-diatom
transitions are forbidden in the LLMG model.

The integrability of the LLMG model results from its high internal
symmetry. This symmetry can be violated in real physical systems, e.
 g.,
when Feshbach resonance is used to tune the elastic scattering
 length. The
effect of Feshbach resonance appears in 2B scattering when the
 collision
energy of a pair of atoms in an open channel is close to the energy
 of a
bound state (resonant molecule) in a closed channel (see Ref.\
\cite{TTHK99}). Two-body Feshbach resonance collisions in atomic
waveguides have been analyzed in Ref.\ \cite{Y05}. The results
 demonstrate
that when the collision energy is low enough compared to the trap
frequency, the problem can be treated as essentially 1D, but with an
energy-dependent zero-range interaction. The present work demonstrates
that a Feshbach resonance leads to non-integrability of few-body
 problems,
allowing for chemical processes (association and dissociation), as
 well as
atom-diatom reflection.

The approach used in Ref.\ \cite{Y05} for two-atom scattering leads in
a 3B problem to a system of two coupled equations for the
 wavefunctions in
the momentum representation, $\varphi _{0}\left(
 q_{1},q_{2},q_{3}\right) $ and $\varphi _{1}\left( q_{1},q_{m}\right
) $, associated with
the three-atom and atom-molecule channels, respectively. Using a
 system of
units in which $\hbar =1$, the equations are
\begin{eqnarray}
E \varphi _{0}\left( q_{1},q_{2},q_{3}\right) ={1\over
 2m}\sum\limits^{3}_{j=1}q^{2}_{j}\varphi _{0}\left(
 q_{1},q_{2},q_{3}\right)  \nonumber
\\
+{1\over 2\pi }U_{a}\sum\limits^{3}_{j=1}\int d^{3}q^\prime \delta
 \left( q^\prime _{j}-q_{j}\right) \delta \left( Q-Q^\prime \right)
 \varphi _{0}\left( q_{1}^\prime ,q_{2}^\prime ,q_{3}^\prime \right)
 \nonumber
\\
+\left( 3\pi \right) ^{-1/2}g^{*}\sum\limits^{3}_{j=1}\varphi
 _{1}\left( q_{j},Q-q_{j}\right)
\\
E\varphi _{1}\left( q_{1},q_{m}\right) =\left( {q{ } ^{2}_{1}\over
 2m}+{q{ } ^{2}_{m}\over 4m}+D_{1D}\right) \varphi _{1}\left(
 q_{1},q_{m}\right)  \nonumber
\\
+\left( {3\over \pi }\right) ^{1/2}g\int dq_{3}\varphi _{0}\left(
 q_{1},q_{m}-q_{3},q_{3}\right)  . \label{phi1}
\end{eqnarray}
Here $q_{j}$  are the momenta of the atoms, $q_{m}$  is the momentum
 of the
resonant molecule, and $Q=q_{1}+q_{2}+q_{3}$  is the center-of-mass
 momentum. The
parameters of the 1D problem --- the atom-atom interaction strength
 $U_{a}$, the
atom-molecule coupling strength $g$, and the detuning between the
 atomic and
molecular states $D_{1D}$--- can be expressed in terms of 3D
 resonance scattering
parameters (see Ref.\ \cite{Y05}). The indistinguishability of the
 bosonic
atoms leads to the symmetry of the wavefunction $\varphi _{0}\left(
 q_{1},q_{2},q_{3}\right) $ over permutation
of the atomic momenta. The energy $E$ is counted from the threshold
 of the
three-atom channel.

Equation (\ref{phi1}) neglects the interaction associated with
collisions between the resonant molecule and the third atom. It
 allows a
simple elimination of the atom-molecule channel function $\varphi
 _{1}$. As a result,
the three-atom channel wavefunction obeys the equation
\begin{eqnarray}
E \varphi _{0}\left( q_{1},q_{2},q_{3}\right) ={1\over
 2m}\sum\limits^{3}_{j=1}q^{2}_{j}\varphi _{0}\left(
 q_{1},q_{2},q_{3}\right)  \nonumber
\\
+{1\over 2\pi }\sum\limits^{3}_{j=1}U_{\text{eff}}\left( q_{j}\right)
  \int d^{3}q^\prime \delta \left( q^\prime _{j}-q_{j}\right) \delta
 \left( Q-Q^\prime \right) \varphi _{0}\left( q_{1}^\prime
 ,q_{2}^\prime ,q_{3}^\prime \right)  ,
\end{eqnarray}
involving an effective atom-atom interaction strength
\begin{equation}
U_{\text{eff}}\left( q\right) =U_{a}+{2|g|{ } ^{2}\over E
+i0-D_{1D}-3q^{2}/\left( 4m\right) } . \label{Ueff}
\end{equation}
This strength appears to be the same function of the collision
 momentum
as in the 2B problem (see Ref.\ \cite{Y05}), but here the collision
 energy of
two atoms is expressed in terms of the total energy $E$ and the third
 atom
momentum $q$ due to momentum and energy conservation (letting $Q=0$
 in the
center-of-mass system used here and below).

For a momentum dependent $U_{\text{eff}}$  the internal symmetry of
 the LLMG model
is broken, and a resonant case does not allow a Bethe-ansatz
 solution. This
case is analyzed here by a numerical solution of the Faddeev-Lovelace
equations. Using the conventional Faddeev reduction method (see Ref.\
\cite{Glockle}), one obtains a 1D integral equation
\begin{eqnarray}
X\left( p,p_{0}\right) =2Z\left( p,p_{0}\right) +{m{ } ^{2}\over
 2\kappa { } ^{3}}\int dq Z\left( p,q\right) T_{1D}\left( q\right)
 X\left( q,p_{0}\right)  \nonumber
\\
Z\left( p,q\right) ={2\kappa { } ^{3}\over \pi m}{1\over mE
+i0-p^{2}-pq-q{ } ^{2}} \label{FadLov}
\end{eqnarray}
for the symmetric transition amplitude $X\left( p,p_{0}\right) $.
 Here the 2B scattering
amplitude $T_{1D}\left( q\right) $, corresponding to the
 momentum-dependent $U_{\text{eff}}\left( q\right) $, has the
form (see Ref.\ \cite{Y05})
\begin{equation}
T_{1D}\left( q\right) =U_{\text{eff}}\left( q\right) \left\lbrack 1
+{i\over 2}mU_{\text{eff}}\left( q\right) \left( mE-{3\over 4}q^{2}
+i0\right) ^{-1/2}\right\rbrack ^{-1}. \label{T1D}
\end{equation}
Equations similar to Eq.\ (\ref{FadLov}) have been used in Refs.\
\cite{Dodd70,Dodd72} for non-resonant 1D problems.

The poles of $T_{1D}\left( q\right) $ at $q=\pm 2\sqrt{m\left(
 E-E_{b}\right) }/3$ correspond to 2B bound states
(diatoms) with the binding energy $E_{b}=\kappa ^{2}/m$. The
 reciprocal diatom size $\kappa $ here
and in Eq.\ (\ref{FadLov}) is a solution of the cubic equation (see
 Refs.\
\cite{Y05,KD98})
\begin{equation}
\kappa ^{3}+{m\over 2}U_{a}\kappa ^{2}+mD_{1D}\kappa +{1\over
 2}m^{2}D_{1D}U_{a}-m^{2}|g|^{2}=0 . \label{kappa}
\end{equation}
These states are superpositions of the closed and open channels. The
probability to find the diatom in the open channel is
$W_{0}=\left( \kappa ^{2}+mD_{1D}\right) /\left( 3\kappa ^{2}
+mU_{a}\kappa +mD_{1D}\right) $. This probability varies from 1 far
 off
resonance to ${1\over 3}$ in resonance for $U_{a}=0$.

In the numerical solution of Eq.\ (\ref{FadLov}) the transition
amplitude is expressed in terms of odd and even amplitudes
$X_{\pm }\left( p,p_{0}\right) =X\left( p,p_{0}\right) \pm X\left(
 -p,p_{0}\right) $, $p>0$, that satisfy uncoupled equations. The
contributions of the poles of $T_{1D}\left( q\right) $ are separated
 from the Cauchy
principal-value integrals as in Ref.\ \cite{Dodd72}.  At $E>0$, when
 chemical
processes are allowed, the function $Z\left( q,p\right) $ has
 singularities on the real
axis too. Unlike the 3D case with logarithmic singularities, in the
 1D case
the singularities are simple poles and are treated in a similar way
 as the
poles of $T_{1D}\left( q\right) $.

The reflection, transmission, and dissociation probabilities in the
collision of an atom with momentum $p_{0}$  and a diatom with
 momentum $-p_{0}$  can be
expressed in terms of the amplitude $X\left( p,p_{0}\right) $ as,
 respectively,
\begin{eqnarray}
P_{\text{ref}}\left( p_{0}\right) =| {4\pi m\over 3p{ }
 _{0}}W_{0}X\left( -p_{0},p_{0}\right) |^{2} \nonumber
\\
P_{\text{tran}}\left( p_{0}\right) =|1+i{4\pi m\over 3p{ }
 _{0}}W_{0}X\left( p_{0},p_{0}\right) |^{2}
\\
P_{\text{diss}}=1-P_{\text{ref}}-P_{\text{tran}} . \nonumber
\end{eqnarray}
Here the factor $W_{0}$  describes the open-channel fraction in the
 incoming
and outgoing waves. The results can be conventionally expressed in
 terms of
dimensionless parameters: the non-resonant interaction strength
$u=m^{1/3}|g|^{-2/3}U_{a}$, the collision energy $\epsilon
 _{2}=3p^{2}_{0}/\left( 4mD_{0}\right) $, and the detuning
$b=D_{1D}/D_{0}$, where an energy scale $D_{0}=m^{1/3}|g|^{4/3}$  is
 used.

\begin{figure}
\includegraphics[width=3.375in]{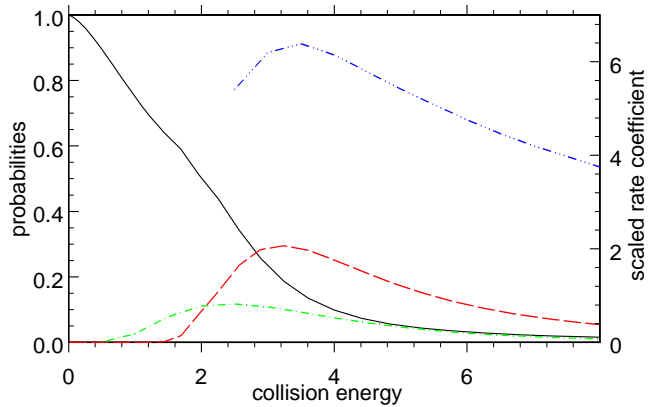}

\caption{Probabilities of reflection (solid line) and
dissociation (dashed line) in atom-diatom collisions calculated for
$b=0$ and $u=0$ as functions of the dimensionless collision energy
 $\epsilon _{2}$. The
scaled rate coefficient of 3B association $K_{3}m/\hbar $, calculated
 as a
function of the three-atom energy $\epsilon _{3}$,  is presented by
 the dot-dashed
($u=0$, $b=0$, $\theta =\pi /6$) and dot-dot-dot-dashed ($u=1.6$,
 $b=1$, $\theta =\pi /10$) lines.}
\label{fig_prob_p}

\end{figure}

Consider an association of three atoms with momenta $p_{1}$, $p_{2}$,
 and $p_{3}$
($p_{1}+p_{2}+p_{3}=0$ in the center-of-mass system). The momenta of
 the resulting
atom and diatom $\pm p_{*}$  are determined by energy conservation, as
\begin{equation}
E={1\over 2m}\left( p^{2}_{1}+p^{2}_{2}+p^{2}_{3}\right) ={3\over
 4m}p^{2}_{*}-E_{b} .
\end{equation}
The association rate coefficient is given by
\begin{eqnarray}
K_{3}\left( p_{1},p_{2},p_{2}\right) ={2\pi ^{2}m{ } ^{3}\over
 27\kappa ^{3}p{ } _{*}}W_{0}\biggl\lbrack
|\sum\limits^{3}_{j=1}T_{1D}\left( p_{j}\right) X\left( p_{j},p_{
*}\right) |^{2} \nonumber
\\
+|\sum\limits^{3}_{j=1}T_{1D}\left( p_{j}\right) X\left( p_{j},-p_{
*}\right) |^{2}\biggr\rbrack  .
\end{eqnarray}
Except for the dimensionless three-atom collision energy $\epsilon
 _{3}=E/D_{0}$, the
rate coefficient depends on the energy distribution between the three
atoms. This distribution can be parameterized by an angle $0\le
 \theta \le \pi /6$, such
that $p_{1}=\sqrt{3mE/4}\cos \theta $, $p_{2,3}=\sqrt{3mE/4}\cos\left
( \theta \pm 2\pi /3\right) $. At $\theta =0$ one of the momenta
attains the maximal possible value $p_{1}=\sqrt{3mE/4}$, while
 $p_{2,3}=-p_{1}/2$. A symmetric
configuration ($p_{1}=-p_{2}=\sqrt{mE}$, $p_{3}=0$) corresponds to
 $\theta =\pi /6$. The association rate
coefficient is plotted below in units of $\hbar /m$.

\begin{figure}
\includegraphics[width=3.375in]{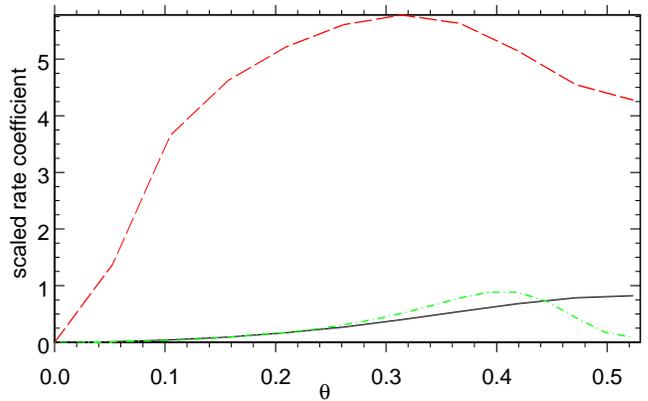}

\caption{The scaled rate coefficient of 3B association $K_{3}m/\hbar
 $ calculated
as a function of $\theta $  for $\epsilon _{3}=2.4$, $u=0$, $b=0$
(solid line), $\epsilon _{3}=4.5$, $b=1$, $u=1.6$
(dashed line), and $\epsilon _{3}=4.5$, $b=1$, $u=-2$ (dot-dashed
 line).} \label{fig_prob_t}

\end{figure}

The calculated probabilities and rate coefficient are presented in
Fig.\ \ref{fig_prob_p} as functions of the corresponding dimensionless
collision energies. The results demonstrate total reflection for slow
collisions. Similar reflection behavior has been obtained in Ref.\
\cite{Dodd72} for a case of non-identical atoms with
 energy-independent
interaction strengths. The dissociation appears above the threshold,
 where
$3p^{2}_{0}/\left( 4m\right) >E_{b}$  (or $\epsilon _{2}>1$ for the
 conditions in Fig.\ \ref{fig_prob_p}). The 3B
association rate attains its maximum at a collision energy $\epsilon
 _{3}$  ranging
between 2 and 4 (depending on $u$, $b$, and $\theta $), and decreases
 to zero for slow
collisions. These processes vanish in fast collisions, when the
interactions become small perturbations. The dependence of the
 association
rate on the energy distribution between the atoms behaves differently
 for
various $u$ and $b$ values (see Fig.\ \ref{fig_prob_t}). No
 association occurs
at $\theta =0$ when two atoms have equal momenta.

\begin{figure}
\includegraphics[width=3.375in]{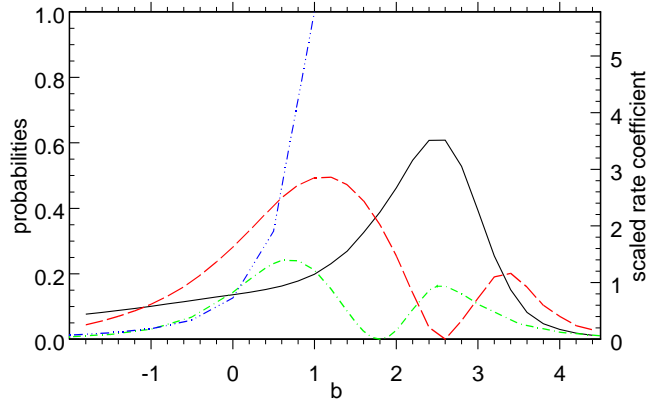}

\caption{Probabilities of reflection (solid line) and dissociation
(dashed line) in atom-diatom collisions calculated for the fixed
dimensionless collision energy $\epsilon _{2}=3.6$ and the detuning
 $b=0$ as functions of
the dimensionless detuning $b$. The scaled rate coefficient of 3B
association $K_{3}m/\hbar $ is presented by the dot-dashed ($u=0$,
 $\epsilon _{3}=2.4$, $\theta =\pi /6$) and
dot-dot-dot-dashed ($u=1.6$, $\epsilon _{3}=4.5$, $\theta =\pi /10$)
 lines.} \label{fig_prob_b}

\end{figure}

Given a fixed collision energy, the reflection, dissociation, and
association persist over a detuning interval (see Fig.\
 \ref{fig_prob_b}).
The oscillating behavior of the probabilities results from
 interference of
the odd and even amplitudes $X_{\pm }\left( p,p_{0}\right) $, which
 are analogs of partial waves in
3D scattering. However, unlike the 3D case, where only the $s$-wave
contributes to low-energy scattering due to a centrifugal barrier, in
 the
1D case both odd and even amplitudes have to be taken into account at
 all
energies. All chemical processes, as well as reflection, vanish at
 large
detunings, when the energy dependence of the interaction strength
 $U_{\text{eff}}$
becomes negligible [see Eq.\ (\ref{Ueff})], restoring integrability
 of LLMG
model. For a repulsive non-resonant interaction ($U_{a}>0$) the bound
 state
exists while $D_{1D}<2|g|^{2}/U_{a}$  [see Eq.\ (\ref{kappa})], or
 while $b<2/u$. As
$b\rightarrow 2/u$ the bound state becomes more shallow, leading to a
 drastic increase
of the association rate. The dependence on the non-resonant
 interaction
strength (see Fig.\ \ref{fig_prob_b}) demonstrates a similar behavior
: a
drastic increase of the association rate near the threshold at $u=2
/b$, and a
decrease of all the rates at large absolute values of $u$ due to the
restoration of integrability.

\begin{figure}
\includegraphics[width=3.375in]{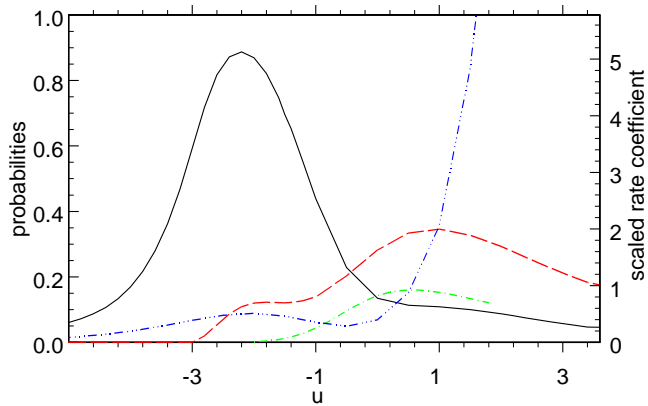}

\caption{Probabilities of reflection (solid line) and dissociation
(dashed line) in atom-diatom collisions calculated for fixed
 dimensionless
collision energy $\epsilon _{2}=3.6$ and detuning $b=0$ as functions
 of the dimensionless
non-resonant interaction strength $u$. The scaled rate coefficient of
 3B
association $K_{3}m/\hbar $ is presented by the dot-dashed ($b=0$,
 $\epsilon _{3}=2.4$, $\theta =\pi /6$) and
dot-dot-dot-dashed ($b=1$, $\epsilon _{3}=4.5$, $\theta =\pi /10$)
 lines.} \label{fig_prob_u}

\end{figure}

Thus, a substantial association rate can be observed for
collision energies ranging over several units  $D_{0}$, which must lie
below the transverse frequency $\omega _{\perp }$, in order to
 conserve the quasi-1D
behavior of the system. This condition is satisfied for relatively
weak resonances $\mu \Delta <\omega _{\perp }a_{\perp }/a_{3D}$  and
 small detunings $\mu \left( B-B_{0}-\Delta \right) -
\omega _{\perp }<\sqrt{\omega _{\perp }\mu \Delta a_{\perp }/a{ }
 _{3D}}$, when the relations between the 1D and 3D scattering
parameters of Ref.\ \cite{Y05} are applicable. These relations lead to
expressions for the energy unit,
\[
D_{0}=m^{1/3}\left( \omega _{\perp }a_{3D}\mu \Delta \right) ^{2
/3}\left( 1-C{a{ } _{3D}\over a{ } _{\perp }}\right) ^{-4/3} ,
 \nonumber
\]
and the dimensionless parameters
\begin{eqnarray}
u=\left( {8m\omega ^{2}_{\perp }a{ } ^{2}_{3D}\over \mu \Delta
 }\right) ^{1/3}\left( 1-C{a{ } _{3D}\over a{ } _{\perp }}\right)
 ^{-1/3} \nonumber
\\
b={1\over D{ } _{0}}\left\lbrack \mu \left( B-B_{0}\right) -\omega
 _{\perp }+C{a{ } _{3D}\over a{ } _{\perp }}\mu \Delta \left( 1-C{a{
 } _{3D}\over a{ } _{\perp }}\right) ^{-1}\right\rbrack  . \nonumber
\end{eqnarray}
Here $\Delta $ is the phenomenological resonance strength, $\mu $ is
 the difference
between the magnetic momenta of an atomic pair in the open and closed
channels,  $B- B_{0}$  is the detuning of the external magnetic field
 $B$ from its
resonant value $B_{0}$,  $a_{\perp }=\left\lbrack 2/\left( m\omega
 _{\perp }\right) \right\rbrack ^{1/2}$  is the transverse harmonic
 oscillator
length, and $C\approx 1.4603$ (see Ref.\ \cite{O98}). For example, in
 a waveguide
with the transverse frequency $\omega _{\perp }=50\times 2\pi $  KHz,
 collisions remain quasi-1D for
the collision energy less then $2.4 \mu $K. In this case $D_{0}=0.27
 \mu $K, $u=0.22$,
$dB/db=1.1$  mG, and $\hbar /m=2.6\times 10^{-5}$  cm$^{2}/$s for the
 Na resonance at 853 G;
$D_{0}=0.48 \mu $K, $u=0.53$, $dB/db=0.52$  mG, and $\hbar
/m=6.9\times 10^{-6}$  cm$^{2}/$s for the $^{87}$Rb
resonance at 685 G; and $D_{0}=0.35 \mu $K, $u=1.7$, $dB/db=9.3$  mG,
 and $\hbar /m=4.5\times 10^{-6}$
cm$^{2}/$s for the Cs resonance at 20 G.

The high internal symmetry of the LLMG model can be also violated due
to virtual transverse mode excitation even if the interaction of free
atoms is non-resonant. This effect has been considered in Ref.\
\cite{MEG05} for 3B collisions in atomic waveguides and in Ref.\
\cite{SCKB05} for atom-soliton collisions. It can be also analyzed
 using
the present approach, since a confinement-induced resonance can be
represented as a kind of Feshbach resonance (see Ref.\ \cite{BMO03}).
 The
scattering amplitude in an atomic waveguide (see Eq.\ (6.9) in Ref.\
\cite{MBO04}) can be approximated by Eq.\ (\ref{T1D}) with the proper
choice of the parameters $g$, $D_{1D}$, and $U_{a}$. However, now the
 problem contains
the single energy scale, $\omega _{\perp }$, and the energy unit
 $D_{0}$  will have the same
order. As a result, the association and dissociation rates are
 negligibly
small in the quasi-1D regime and the only evidence of nonintegrability
remains the atom-diatom reflection, in agreement with Refs.\
\cite{MEG05,SCKB05}. A narrow Feshbach resonance can provide reacher
physical phenomena. In addition to association, dissociation, and
reflection the nonintegrability can lead to a change of the asymptotic
momentum set in three-atom elastic collisions.

In conclusion, the 3B problem with two-channel 2B interactions can be
reduced to a single-channel case with an energy-dependent interaction
strength. The energy dependence violates the high internal symmetry
 of the
LLMG model, allowing chemical reactions. Reflection becomes the
 dominant
output channel of atom-diatom low-energy collisions, while
 dissociation
appears above a threshold. The 3B association should be observable in
experiments.


\begin{thebibliography}{99}
\bibitem{G01a} A. G\"orlitz {\it et al.}, Phys. Rev. Lett. {\bf 87}
130402 (2001).
\bibitem{S02} K. Strecker {\it et al.}, Nature {\bf 417}, 150 (2002).
\bibitem{K02} L. Khaykovich {\it et al.}, Science {\bf 296}, 1290
(2002).
\bibitem{R03}  S. Richard  {\it et al.}, Phys. Rev. Lett. {\bf 91}
010405 (2003); M. Hugbart  {\it et al.}, physics/0501456.
\bibitem{G01} M. Greiner {\it et al.}, Phys. Rev. Lett. {\bf 87,}
160405 (2001); H. Moritz  {\it et al.}, Phys. Rev. Lett. {\bf 94},
 210401
(2005).
\bibitem{W04}T. Kinoshita {\it et al.}, Science {\bf 305}, 1125
(2004).
\bibitem{F05} C. D. Fertig  {\it et al.}, Phys. Rev. Lett. {\bf 94},
120403 (2005).
\bibitem{L02} A. E. Leanhardt {\it et al.}, Phys. Rev. Lett. {\bf 89,}
040401 (2002).
\bibitem{F02}R. Folman {\it et al.}, Adv. At. Mol. Opt. Phys. {\bf 8},
263 (2002).
\bibitem{S05}Y. Shin {\it et al.}, cond-mat/0506464.
\bibitem{O98} M. Olshanii, Phys. Rev. Lett. {\bf 81,} 938 (1998).
\bibitem{BMO03} T. Bergeman, M. Moore, and M. Olshanii,  Phys. Rev.
Lett. {\bf 91,} 163201 (2003).
\bibitem{MBO04} M. Moore, T. Bergeman, and M. Olshanii, J. Phys.
(Paris) IV {\bf 116}, 69 (2004).

\bibitem{TTHK99}E. Timmermans, P. Tommasini, M. Hussein, and A.
Kerman, Phys. Rep. {\bf 315}, 199 (1999).

\bibitem{Y05}V. A. Yurovsky,  Phys. Rev. A {\bf 71}, 012709 (2005).

\bibitem{LL63} E. H. Lieb and W. Liniger, Phys. Rev. {\bf 130}, 1605
(1963).
\bibitem{McGuire64}J. B. McGuire, J. Math. Phys. {\bf 5}, 622 (1964).

\bibitem{Glockle} W. Glockle, {\it The Quantum Mechanical Few-Body
Problem} (Springer, Berlin, 1983).

\bibitem{Dodd70} L. R. Dodd, J. Math. Phys. {\bf 11}, 207 (1970); C.
K. Majumdar, J. Math. Phys. {\bf 13}, 705 (1972).
\bibitem{Dodd72} L. R. Dodd, Aust. J. Phys. {\bf 25}, 507 (1972).

\bibitem{KD98}K. V. Kheruntsyan and P. D. Drummond, Phys. Rev. A {\bf
 58},
2488 (1998).

\bibitem{MEG05}C. Mora, R. Egger, and A. O. Gogolin, Phys. Rev.
A {\bf 71}, 052705 (2005).

\bibitem{SCKB05}S. Sinha, A. Yu. Cherny, D. Kovrizhin, and J. Brand,
cond-mat/0506060.

\end{thebibliography}
\end{document}